\documentclass{elsart1p}
\usepackage{graphicx}
\usepackage{epstopdf}
\usepackage{amssymb}
\begin{document}
\begin{frontmatter}
%
%
%
%
%
\title{Probing the QCD Critical Point by Higher Moments of Net-Charge Distribution }
%
 \author{Nihar Ranjan Sahoo (for the STAR Collaboration)}
\address{Variable Energy Cyclotron Centre,1/AF Bidhan Nagar,Kolkata - 700064}

%



\begin{abstract}
The Beam Energy Scan program has been undertaken at the Relativistic
Heavy Ion Collider (RHIC) to search for the QCD critical point. The
presence of the critical point is expected to lead to non-monotonic
behavior of several quantities. Here we report the result of higher
moments of net-charge distributions for Au+Au collisions at
 $\sqrt{s_{NN}}$
 = 39 GeV 
as measured by the STAR experiment. The STAR results are compared with results from HIJING event generator and Hadron Resonance (HRG) Models.

\end{abstract}

%
%

\end{frontmatter}

\section{Introduction}
\label{}
Lattice calculations[1] have predicted the presence of a critical point in the phase boundary of hadronic matter with Quark-Gluon Plasma (QGP). The critical point is the end point of the first order phase transition. To map the QCD phase diagram and to  locate the critical point, the Beam  Energy Scan program has started in the year 2010 at the Relativistic Heavy-Ion Collider at Brookhaven National Laboratory. The program aims to scan the beam energy from  $\sqrt {s_{NN}}$ = 5 GeV to 200 GeV corresponding to baryon chemical potentials up to as high as 550 MeV and down to 14 MeV[2]. The characteristic signature for the CP is the divergence of the susceptibilities of conserved quantities like net-charge, net-baryon, net-strangeness[4] and the correlation length($\xi$)[1,3]. These quantities are related to the higher moments of the event-by-event distribution of the above conserved quantities. The variance, skewness and kurtosis are related to $\xi^{2}$, $\xi^{4.5}$ and $\xi^{7}$[3] respectively. The ratio of the fourth to the second order susceptibilities are related to the product of the kurtosis and variance($K\sigma^{2}$). It is expected that these higher moments will be non-monotonic, at the CP,  as a function of an experimentally varied parameter such as the beam energy[5]. 
     
     We present  the preliminary results of higher moments  and their product of net-charge distribution from Au+Au collision at 39 GeV and 200 GeV measured by the STAR experiment at the RHIC.

\section{Analysis Details and Results}
\label{}
The STAR experiment provides the excellent particle identification and large acceptance for event-by-event fluctuation analysis. The  Time Projection Chamber(TPC) is the main tracking device. It is used to identify the charge particles within full azimuthal and  $\pm 1.8 $ unit of pseudo-rapidity($\eta$). The event-by-event net-charge is measured  at $\sqrt{s_{NN}}$ = 39 GeV for Au +Au collisions occurring within 30 cm of the TPC center along the beam line. For this analysis, nearly $6 \times 10^{6}$ minimum bias events are used. The charged particles between  the range $ 0.15 <  p_{T} < 1.0 $ GeV/c are measured at $|\eta| < 0.5 $ region. Standard STAR track quality cuts are used for this analysis. For the electron and positron identification in same $p_{T}$ range, which is a source of the background for the analysis, Time Of Flight(TOF) detector is used. For the centrality selection, uncorrected charged particles multiplicity measured within $|\eta| < 0.5$  from the TPC is utilized . To get the average number of participant ($<N_{part}>$) for each centrality, Glauber model calculation is done. 
  
  \begin{figure}[hb]
   \centering
      \includegraphics[width=75mm]{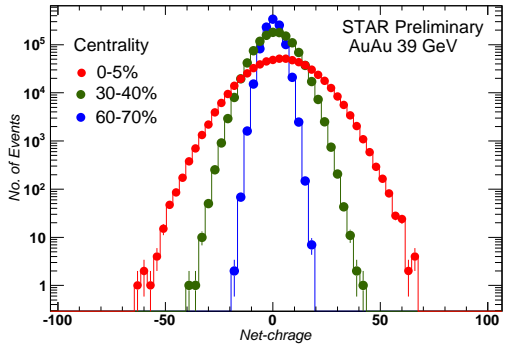} 
   \caption{Net-charge distribution for Au+Au collision at
     $\sqrt{s_{NN}}$  = 39 GeV for 0-5\%, 30-40\% and 60-70\% centrality .}
   \label{fig:example}
\end{figure}

\begin{figure}[hb]
\begin{center}
   \includegraphics[width=90 mm]{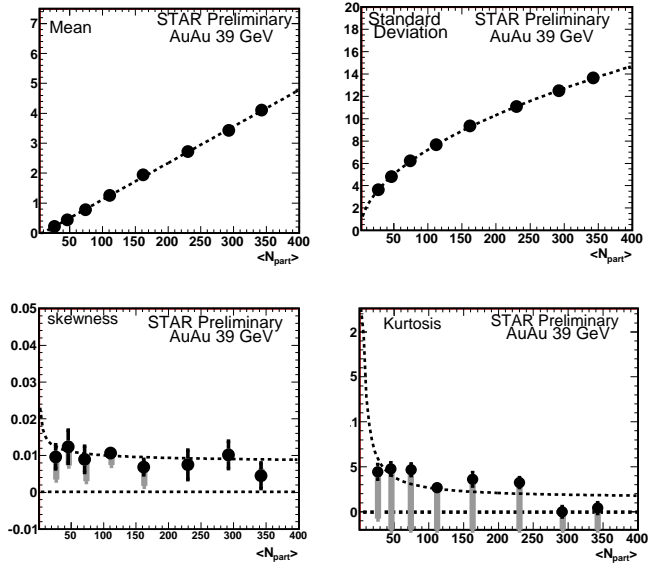}
\caption{The mean(top left), standard deviation(top right), skewness(bottom left) and kurtosis(bottom right) are plotted versus $<N_{part}>$. The dotted lines are the expected  values from the central limit theorem. The error bars are statistical and the grey lines represent  systematic error  due to the centrality bin width effect.   }

\label{default}
\end{center}
\end{figure}

  

\subparagraph*{}
Figure 1 shows the net-charge distribution for centrality 0-5\%, 30-40\% and 60-70\%. The net-charge distributions are symmetric around the mean value for different centralities (small skewness values as seen in Fig. 2).  The HIJING event generator is used for model comparison with data in same $\eta$ and $p_{T}$ region. To understand the leptonic contribution to the higher moments of net-charge distribution, both the model and the data based approaches (utilizing the TOF detector) are made. Both these studies show their contribution is negligible.
 
    The mean, standard deviation, skewness and kurtosis of Au+Au
    collisions at 39 GeV are plotted as a function of $<N_{part}>$
    representing different centrality in Fig. 2. These moments are
    fitted with the expected form of the central limit theorem(CLT) to
    understand the volume dependence of the higher moments[6]. Both
    the mean and standard deviation increase with $<N_{part}>$, where
    the mean shows linear dependence. The error bars include both
    statistical errors and systematic errors due to the range of
    multiplicity within each centrality bin. Reduction of the centrality bin width effect and the other systematics studies are in progress. In more central collisions, the kurtosis shows a deviation from the CLT that need more study to understand this effect.
    
  \begin{figure}[htbp]
\begin{center}
\includegraphics[width=58.5mm]{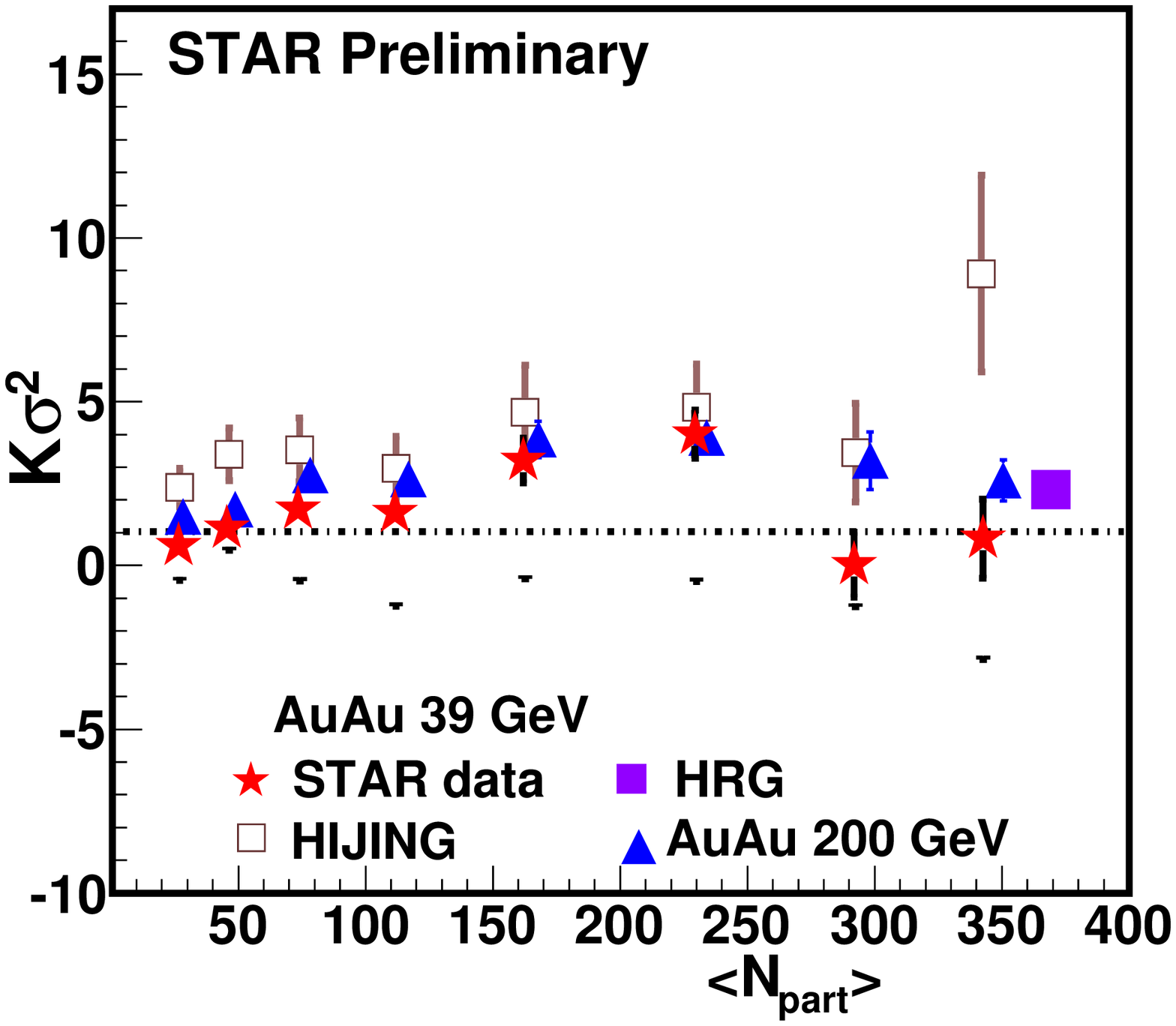}
\includegraphics[width=58mm]{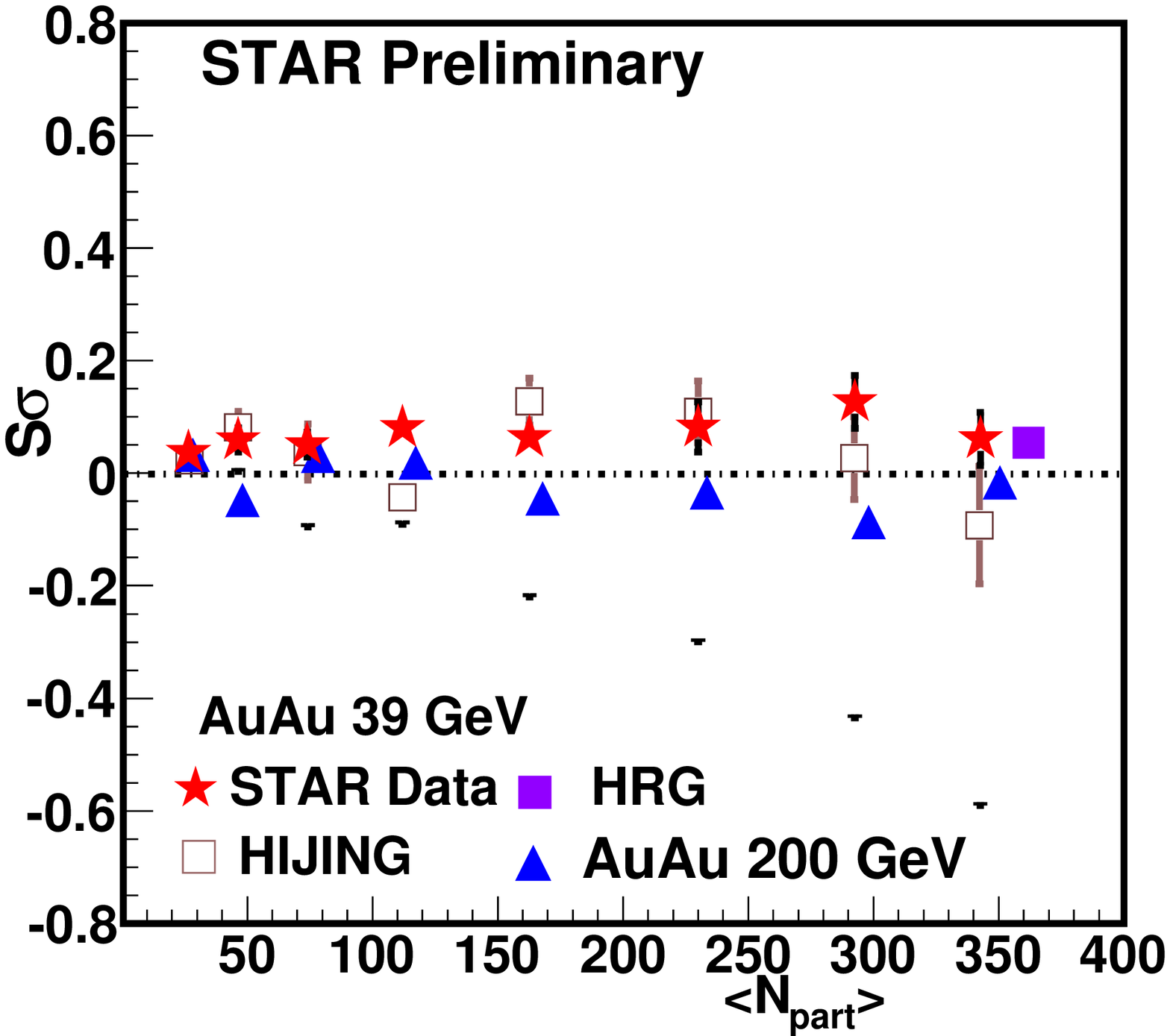}
\caption{The $K\sigma^{2}$ (left-panel) and $S\sigma$ (right panel) of the net-charge distribution are plotted versus $<N_{part}>$ respectively from the data. The filled rectangle represent the HRG model results from [8]. The blank rectangles are the HIJING model results for the Au+Au at 39 GeV. The red star and the blue triangle represent Au+Au 39 and 200 GeV results from the data.  The error bars are statistical error and black caps represent  systematic error due to the centrality bin width effect. }
\label{default}
\end{center}
\end{figure}

To eliminate the volume dependence in moments, the product of the
kurtosis and variance ($K\sigma^{2}$) and that of the skewness and
standard deviation ($S\sigma$) are used. $K\sigma^{2}$ and $S\sigma$
are plotted versus $<N_{part}>$ in Fig. 3 for Au+Au collisions at 39
GeV and 200 GeV. Here Au+Au 200 GeV results are taken from [7]. The
results from the Hadron Resonance Gas(HRG) model[8] are consistent
with the data within the errors. The results from HIJING are also
agreement with the data. The $K\sigma^{2}$ values are larger than
unity where the $S\sigma$ values are close to zero for different
centralities. The $K\sigma^{2}$ and $S\sigma$ values are similar for
Au+Au collisions at  $\sqrt{s_{NN}}$  = 39 and 200 GeV.

\section{Summary}
\label{}
 The STAR preliminary results of the higher moments of the net-charge
 distribution for Au+Au collision at  $\sqrt{s_{NN}}$  = 39 and 200 GeV are presented. The results are compared with the HRG and HIJING models. Within current experimental uncertainties, the $K\sigma^{2}$ and $S\sigma$ values are comparable for collision at 39 and 200 GeV. The results are in agreement with the HRG model. The results from HIJING model, which does not include CP physics, are consistent with those from data. The data analysis and the model studies are in progress for the other BES energies.


\end{document}